\documentclass[prl,twocolumn,showpacs,superscriptaddress,nofootinbib]{revtex4-1}
\usepackage{graphicx}
\usepackage{bm}
\usepackage{amssymb}
\usepackage{color}
\usepackage{amsmath}
\usepackage{amstext}
\usepackage{latexsym}
\usepackage[usenames,dvipsnames]{xcolor}
\usepackage[colorlinks=true,citecolor=Blue,linkcolor=RubineRed,urlcolor=Blue]{hyperref}

\usepackage{mathptmx} 

\def\la{\langle}
\def\ra{\rangle}

\def\tr{\rm{Tr}}

\newcommand{\beq}{\begin{equation}}
\newcommand{\eeq}{\end{equation}}
\newcommand{\beqa}{\begin{eqnarray}}
\newcommand{\eeqa}{\end{eqnarray}}

\begin{document}

\title{Universal Statistics of Topological Defects Formed in a  Quantum Phase Transition}
\author{Adolfo del Campo}
\affiliation{Department of Physics, University of Massachusetts, Boston, MA 02125, USA}
\affiliation{Theoretical Division, Los Alamos National Laboratory, Los Alamos, NM 87545, USA}

\begin{abstract}
When a quantum phase transition is crossed in finite time,  critical slowing down leads to the breakdown of adiabatic dynamics and the formation of topological defects. The average density of defects scales with the quench rate following a universal power-law predicted by the Kibble-Zurek mechanism. We analyze the  full counting statistics of kinks and report the exact  kink number distribution  in the transverse-field quantum Ising model. Kink statistics is described by the Poisson binomial distribution with all cumulants exhibiting a universal power-law scaling with the quench rate. In the absence of finite-size effects, the distribution approaches a normal one, a feature that is expected to apply broadly in systems described by the Kibble-Zurek mechanism.

\end{abstract}

\maketitle

Across a quantum phase transition, the equilibrium relaxation time diverges. This phenomenon, known as critical slowing down, is responsible for  the nonadiabatic character of critical dynamics. Preparing the ground state of the broken-symmetry phase, an ubiquitous task in quantum science and technology,  is thus intrinsically challenging: traversing the phase transition in finite time leads to the formation of topological defects. 
The Kibble-Zurek mechanism (KZM) is the paradigmatic theory to describe this scenario \cite{Dziarmaga10,Polkovnikov11,DZ14}.
Its origins are found in the pioneering insight by Kibble on the role of causality in structure formation in the early universe \cite{Kibble76a,Kibble76b}.
Soon after, it was pointed out by Zurek that condensed-matter systems offer a test-bed to study the dynamics of symmetry breaking \cite{Zurek96a,Zurek96b,Zurek96c}. The key prediction of the KZM is that the average density $d$ of the resulting topological defects scales with the quench time $\tau_Q$ in  which the phase transition is crossed as a universal power-law, $d\propto\tau_Q^{-\alpha}$.  The power-law exponent $\alpha=D\nu/(1+\nu z)$ is set by a combination of the dimensionality of the system $D$, and the dynamic and correlation-length (equilibrium) critical exponents denoted by $z$ and $\nu$, respectively. 

The validity of the KZM is however not restricted to the classical domain.
The paradigmatic Landau-Zener formula, describing excitation formation in two-level systems, was shown to capture the KZM for long quench times \cite{Damski05,ZDZ05}. As a result, paradigmatic models exhibiting quantum phase transitions, such as the 1D Ising chain, could be shown to obey the KZM, establishing the validity of the mechanism in the quantum domain for thermally isolated systems \cite{Polkovnikov05,Damski05,Dziarmaga05,ZDZ05}. 
Due to its broad applicability, the KZM stands out as a result in  statistical mechanics describing nonequilibrium properties (density of defects) in terms of equilibrium quantities (critical exponents). On the applied side, it suggest the need to pursue adiabatic strategies in quantum simulations as well as in  quantum annealing, where the mechanism provides useful heuristics.

Under unitary dynamics the state of the system following the crossing of the phase transition is characterized by collective and coherent quantum excitations.
One can thus expect that even for isolated quantum systems, the order parameter in the broken symmetry phase as well as the number of topological defects exhibit fluctuations and are characterized by a probability distribution. In the classical domain, the study of the equilibrium probability distribution of the order parameter has proved useful in spin systems \cite{Bruce81,Binder1981,Nicolaides88}, and it is known to be universal in the scaling limit \cite{Binder1981,Aji01}. In the quantum domain, progress has been made by analyzing the equilibrium distribution of the magnetization in a variety of critical spin systems  \cite{Lamacraft08,DeChiara16} or following a sudden quench \cite{Groha18}.

Studies of the distribution of topological defects generated in the course of a phase transition have been limited to winding numbers. In both classical and quantum systems, the distribution is known to have zero mean value  and a dispersion typical of a random walk with a number of steps that can be estimated with the KZM \cite{Uwe07,Uwe10,Uhlmann10,Das12,Sonner15}.  

In this Letter, we consider the critical dynamics of the one-dimensional quantum Ising model in a transverse field  and analyze the distribution of topological defects formed during the crossing of the critical point in finite time. 
The mean of the kink number distribution reproduces the prediction by the KZM, as expected. We focus on the characterization of the fluctuations of the kink number distribution and show that all higher order cumulants share the universal power-law scaling with the quench time in which the phase transition is traversed. Our results thus show that the nonadiabatic dynamics leading to the formation of topological defects exhibits a universal behavior beyond the scope of the Kibble-Zurek mechanism, that  determines the average density of defects. Said differently, the KZM can be extended to account for the full  distribution  of topological defects.


{\it The quantum Ising model in a transverse field.---}
As a paradigmatic model of a quantum phase transition we consider the one dimensional quantum Ising model \cite{Sachdev,Chakrabarti96}.  The Hamiltonian of a chain of $N$ spins in a transverse magnetic field $g$ reads,
\beqa
\mathcal{H}=-J\sum_{m=1}^N(\sigma_m^z\sigma_{m+1}^z+g \sigma_m^x).
\label{H_Ising}
\eeqa
Its experimental study is  amenable  via quantum simulation that has been reported in a variety of platforms including  
trapped ions, \cite{Monroe11,Monroe17}, superconducting circuits \cite{Johnson12,Barends2016a}, Rydberg gases \cite{Labuhn2016}, and NMR experiments \cite{Li14}.
We consider periodic boundary conditions $\sigma_{N+1}=\sigma_{1}$ with even  $N$, for simplicity. The phase diagram of the system is characterized by two  critical points $g_c = \pm 1$ separating a paramagnetic phase 
 ($|g|>1$) and ferromagnetic phase ($|g|<1$).  

The Hamiltonian (\ref{H_Ising}) can be written as a free fermion model, 
making use of the  Jordan-Wigner transformation,
 $\sigma_m^x=1-2c_m^{\dag}c_m$, $\sigma_m^x=-(c_m+c_m^{\dag)} \prod_{\ell<m}(1-2c_\ell^{\dag}c_\ell)$, where $c_m$ are fermionic annihilation operators.  As $\mathcal{H}$ commutes with the parity operator, we shall focus on the even parity subspace, that includes the ground state of the system. Next we define the Fourier transform $c_m=e^{-i\pi/4} \sum_k c_k e^{ikm} / \sqrt{N}$, where the momenta allowed by the boundary conditions  are $k \in \{\pm \pi /N, \pm 3\pi /N, \ldots, \pm(N-1)\pi /N \} $,  and we take the lattice spacing as a unit of length. As shown in \cite{Dziarmaga05}, see as well \cite{Sachdev,Chakrabarti96} and \cite{SM}, the Ising chain Hamiltonian is then given by
 \begin{eqnarray}
\label{hk2}
\mathcal{H}= 2\sum_{k>0} && \psi_k^\dagger \left[ \sigma_k^z (g-\cos k)+\sigma_k^x \sin k  \right] \psi_k, 
\end{eqnarray}
in terms of the operators $\psi_k^\dagger \equiv (c_k^\dagger, c_{-k})$. 
In this form, it becomes apparent  that the critical dynamics of the Ising model can be described via the dynamics of an ensemble of non-interacting two-level systems \cite{Dziarmaga05}. 

To study the quantum critical dynamics, 
we consider an Ising chain initially prepared in the ground state, deep in the paramagnetic phase. The paramagnet is driven across the phase transition by a time-dependent magnetic field of the form
\beqa
\label{gt}
g(t)=g_c\left(1-\frac{t}{\tau_Q}\right),
\eeqa
where $g_c=1$ and $\tau_Q$ is known as the quench time. 
The closing of the gap as the critical point is approached leads to nonadiabatic dynamics and is responsible for the  formation of topological defects, i.e., kinks in the  the quantum Ising chain. We shall be interested in the  distribution of the number of kinks  in the nonequilibrium state reached upon completion of the phase transition at $t=\tau_Q$. With $g(\tau_Q)=0$  the Hamiltonian  is then that of a pure ferromagnet.

{\it Kink number distribution.---}
The operator measuring  the number of kinks  reads
\beqa
\hat{\mathcal{N}} ~\equiv~ \frac12 \sum_{n=1}^{N} 
                  \left(1-\sigma^z_n\sigma^z_{n+1}\right)
\label{calN}
\eeqa
and commutes with the final Hamiltonian at $t=\tau_Q$.
With it, we can construct the projector onto the subspace with a given number of kinks $n$, which can be conveniently written as
\beqa
\delta[\hat{\mathcal{N}} -n]
=\frac{1}{2\pi}\int_{-\pi}^{\pi}d\theta e^{i\theta (\hat{\mathcal{N}} -n)},
\eeqa
using the integral representation of the Kronecker delta. A similar expression can be used for related observables such as the  distribution of the density of kinks, that takes continuous values, using Dirac's delta function instead.
The kink number distribution is given by the expectation value of this operator
\beqa
P(n)=\left\la \delta[\hat{\mathcal{N}} -n]\right\ra,
\eeqa
where the angular bracket denotes the expectation value with respect to the state of the system.
In what follows, it will prove convenient to introduce its Fourier transform representation
\beqa
\label{pneq}
P(n)=\frac{1}{2\pi}\int_{-\pi}^{\pi}d\theta \widetilde{P}(\theta;\tau_Q) e^{-i\theta n}, 
\eeqa
where the characteristic function $ \widetilde{P}(\theta;\tau_Q)$ reads
\beqa
 \widetilde{P}(\theta;\tau_Q)=\tr\left[\hat{\rho}_{\tau_Q} e^{i\theta \hat{\mathcal{N}}}\right].
\eeqa
This expression, being the moment generating function, contains the exponential of the kink number operator, which is naturally highly-nonlocal in real space.
However, it admits a simple representation in Fourier space, as
\beqa
\hat{\mathcal{N}}=\sum_k \gamma_k^\dagger \gamma_k, 
\eeqa
where $\gamma_k$ are the quasiparticle operators that diagonalize the Hamiltonian (\ref{hk2}), i.e., 
$\mathcal{H}=\sum_kE_k(\gamma_k^\dagger \gamma_k-1/2)$.
In addition, for quasi-free fermions (with periodic boundary conditions), the time-dependent  density matrix preserves the tensor product structure during unitary time-evolution. In particular, upon completion of the protocol, the quantum state of the Ising chain is  given by $\hat{\rho}_{\tau_Q}=\bigotimes_k\hat{\rho}_{k,\tau_Q}$, where  $\hat{\rho}_{k,\tau_Q}$ is  the density matrix  of the $k$-mode. As a result, the characteristic function factorizes as
\beqa
 \widetilde{P}(\theta;\tau_Q)=\prod_k\tr\left[\hat{\rho}_{k,\tau_Q} e^{i\theta  \gamma_k^\dagger \gamma_k}\right], 
\eeqa
i.e., it reduces to the product of the characteristic function for each mode $k$.
Said differently, the study of the probability distribution of the density of defects in an Ising chain is equivalent to the study of the full counting statistics of the number of quasiparticles in each mode. The treatment of the latter resembles early studies in quantum transport in mesoscopic  physics focused on  the counting of electrons \cite{LL93}. 
Using the fact that $\gamma_k^\dagger \gamma_k$ is a Fermion number operator  with  eigenvalues   $\{0,1\}$, one can further simply this expression to find 
\beqa
\label{charfunc}
 \widetilde{P}(\theta;\tau_Q)&=&\prod_k\tr\left[\hat{\rho}_{k,\tau_Q}\left(\mathbb{I}_2 +(e^{i\theta}-1)\gamma_k^\dagger \gamma_k\right)\right]\nonumber\\
&=&\prod_{k}\left[1 +(e^{i\theta}-1)\la \gamma_k^\dagger \gamma_k\ra\right].
\eeqa
We note that Eq. (\ref{charfunc}) is the characteristic function associated  with $N$ independent random Bernouilli variables (one for each mode) each of which can take value $1$ (mode excited) with probability $p_k$ and value $0$ (mode in ground state) with probability $(1-p_k)$. This is precisely the characteristic function of the Poisson binomial distribution.  The latter is expected to account for the full counting statistics of defect formation in quasi-free fermion models in which the number of topological defects is related to the number of quasiparticles. A part from the quantum Ising model, these  include  the XY model in one dimension  as well as the Kitaev  model in one and two  spatial dimensions, among other examples \cite{Sachdev,Chakrabarti96}.

Equation (\ref{charfunc}) is highly advantageous for numerical computations. In addition, it makes possible an analytical treatment.  The dynamics in each mode with a linear ramp of the magnetic field (\ref{gt}) is well-described by the
the Landau-Zener formula that yields \cite{Dziarmaga05}
\beqa
 p_k=\la \gamma_k^\dagger \gamma_k\ra=\exp\left(-\frac{1}{\hbar}2\pi J\tau_Qk^2\right).
\eeqa
In turn, this allows one to compute  the cumulant generating function that is given by
\beqa
& & \log  \widetilde{P}(\theta;\tau_Q)=\sum_k\log\left[1 +(e^{i\theta}-1)\la \gamma_k^\dagger \gamma_k\ra\right]\\
&=&\frac{N}{2\pi}\int_{-\pi}^\pi dk
\log \left[1 +(e^{i\theta}-1)\exp\left(-\frac{1}{\hbar}2\pi J\tau_Qk^2\right)\right],\nonumber
\eeqa
where the last expression holds in the continuum limit. 
We can use the identity
$\log (1+\epsilon)=\sum_{p=1}^\infty(-1)^{p+1}\frac{\epsilon^p}{p}$ and perform the integral over the resulting Gaussian integrand  to find
\beqa
\label{CGFexp}
\log  \widetilde{P}(\theta;\tau_Q)
=-\sum_{p=1}^\infty\frac{(1-e^{i\theta})^p}{p\sqrt{p}}N d\times {\rm erf}\left(\frac{\sqrt{\pi p}}{2d}\right),
\eeqa
where ${\rm erf}(x)$ is the error function and we recognize the mean density of defects
\beqa
d=\frac{1}{2\pi}\sqrt{\frac{\hbar}{2J\tau_Q}},
\eeqa
which was derived in \cite{Dziarmaga05}, validating the KZM in the quantum domain for quasi-free fermion systems.

{\it Scaling limit.---}
In the limit of slow quenches, the  cumulant generating function can be simplified given that the average density predicted by KZM $d\ll1$. To leading order in $1/\tau_Q$ one finds
\beqa
\label{CGFa}
\log  \widetilde{P}(\theta;\tau_Q)=-N d \, {\rm Li}_{3/2}(1-e^{i\theta}),
\eeqa
in terms of the polylogarithmic function ${\rm Li}_{3/2}(x)=\sum_{p=1}^\infty x^p/p^{3/2}$ \cite{Jonquiere1889}. This approximation is equivalent to setting ${\rm erf}\left[\sqrt{\pi p}/(2d)\right]=1$ in Eq. (\ref{CGFexp}). To the best of our knowledge, Eq. (\ref{CGFa}) defines a new probability distribution function $P(n)$.

By definition, the expansion of $\log  \widetilde{P}(\theta;\tau)$ generates the cumulants $\{\kappa_q\}$ of the $P(n)$ distribution according to
\beqa
\log  \widetilde{P}(\theta;\tau_Q)=\sum_{q=1}^\infty \frac{(i\theta)^q}{q!}\kappa_q.
\eeqa
Making use of it, or by direct comparison with (\ref{CGFexp}),  we find
\beqa
\label{cum1}
\kappa_1=\la n \ra = Nd =\frac{N}{2\pi}\sqrt{\frac{\hbar}{2J\tau_Q}},
\eeqa
recovering the result for the mean value dictated by the KZM \cite{Dziarmaga05}.
The variance of the number of kinks, that equals the second cumulant $\kappa_2$, is given by
\beqa
\label{cum2}
\kappa_2=\la n^2\ra-\la n\ra^2=N\frac{2-\sqrt{2}}{4\pi}\sqrt{\frac{\hbar}{2J\tau_Q}}, 
\eeqa
and has the same dependence with the quench rate as the mean density $\la n\ra$, being directly proportional to it. 
Indeed, this conclusion holds for all cumulants of the distribution, which do not vanish, making the kink distribution non-normal. In particular, given the expression for the cumulant generating function (\ref{CGFa}), it is clear that all cumulants are nonzero and proportional to the mean, 
\beqa
\kappa_q\propto \la n\ra =N d
\eeqa
for all integer $q$. 
From Eqs. (\ref{cum1}) and (\ref{cum2}), it follows that $\kappa_2/\kappa_1=(2-\sqrt{2})/2\approx 0.29<1$ showing that the kink statistics is sub-Poissonian, see as well \cite{SM,LeCam60}. 
The third cumulant, that equals the third central moment, is given by $\kappa_3=\la (n-\la n\ra)^3\ra=(1-3/\sqrt{2}+2/\sqrt{3})\la n \ra\approx 0.033 \la n \ra$. Thus, $\kappa_3$ is positive,  indicating that the kink number distribution is slightly leaned to low kink numbers and has a comparatively longer tail at high kink numbers.  

Nonetheless, the  weight of cumulants $\kappa_q$ with $q>2$ relative to the mean quickly approaches zero.
Indeed,  $ {\rm Li}_{3/2}(1-e^{i\theta})\approx-i\theta+3\theta^2/(2\pi^2)$ which is equivalent to set to zero all higher order cumulants.
As shown in \cite{SM}, the kink number distribution can be approximated by a normal distribution with mean 
$\la n\ra=Nd$ and variance $\la n^2\ra-\la n\ra^2=3\la n\ra/\pi^2$, namely,
\beqa
\label{Gausspn}
P(n)&\simeq&\mathbf{N}\left(Nd,\frac{3}{\pi^2}Nd\right)\\
&=&
\frac{1}{\sqrt{6 \la n \ra/\pi}}\exp\left[-\frac{\pi^2(n-\la n\ra)^2}{6\la n\ra}\right],\nonumber
\eeqa
where $\la n\ra$ is given in Eq. (\ref{cum1}) as dictated by the KZM.
Eq. (\ref{Gausspn}) can be understood as a limiting case of the binomial distribution in a sequence of  $Nd/p$ independent trials  in which the probability of forming a kink is $p=1-3/\pi^2\approx 0.69$.   This indicates that  the size of the domains in the broken-symmetry phase can be identified with $\hat{\xi}=p/d$, such that the number of trials is given by  the ratio $N/\hat{\xi}=Nd/p$, which is consistent with previous estimates \cite{Dziarmaga05,ZDZ05}. 
We suggest that the full counting statistics of topological defects in systems obeying KZM is  described by a binomial distribution $B(n,p)$ where the number of domains is set by $N_D=N/\hat{\xi}$ and the  probability for defect formation $p$ is expected to be system dependent.  The probability for $n$ topological defects is then  $P(n)=C_n^{N_D}p^n(1-p)^{N_D-n}$, where $C_n^{N_d}=N_D!/(n!(N_D-n)!)$. For $N_D\gg1$ the distribution becomes normal $P(n)\simeq\mathbf{N}\left(N_Dp,N_Dp(1-p)\right)$, as in (\ref{Gausspn}), with $\kappa_2\propto\kappa_1$.  This prediction is consistent with previous studies on spontaneous currents formation, e.g., in superfluid or superconducting rings \cite{Zurek96a,Zurek96c,Monaco02,Das12,Sonner15,Nigmatullin16}.

{\it Numerical results.---}
%
\begin{figure}[t]
\begin{center}
\includegraphics[width=1\linewidth]{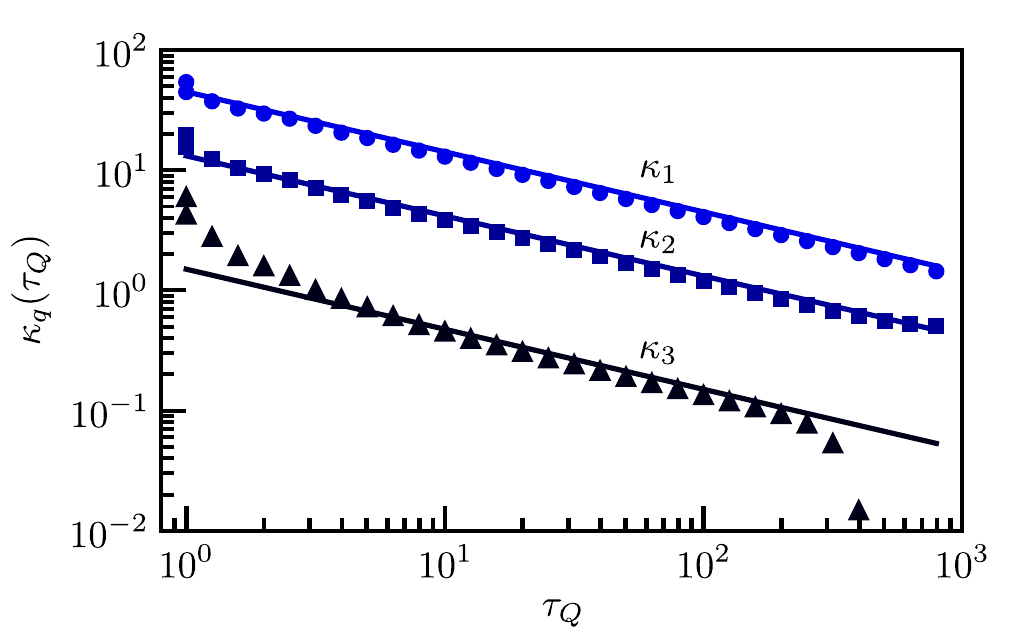}
\end{center}
\caption{\label{fig1}  {\bf Cumulants $\kappa_q$ of the kink number distribution.}  From top to bottom, universal scaling of the mean density of defects ($q=1$), the corresponding variance ($q=2$) and the third cumulant ($q=3$) of the kink number distribution as a function of the quench time $\tau_Q$ in which the phase transition is crossed  ($N=400$). Symbols represent numerical data while solid lines describe the analytical approximation derived in the scaling limit.  The mean density ($q=1$) is predicted by the KZM, see Eq. (\ref{cum1}), and was numerically confirmed in \cite{Dziarmaga05,ZDZ05}. For slow quenches all cumulants exhibit a universal scaling with the  quench time.  The universality of critical dynamics thus extends beyond the scope of the KZM and governs the full distribution of topological defects. Deviations from the scaling limit due to finite-size effects and the onset of adiabatic dynamics are first signaled by high-order cumulants.
}
\end{figure}
To demonstrate the accuracy of these analytical results we perform numerical simulations by integrating the Schr\"odinger equation in Fourier space for each mode. The dynamics of the phase transition is started at $t=-\tau_Q$  and induced by the linear ramp of the magnetic field in Eq. (\ref{gt}). We have checked that the results are robust with respect to other  choices of the initial time $t=-a\tau_Q$ with $a>1$. The final nonequilibrium state is computed at $t=\tau_Q$, deep in the ferromagnetic phase.  Evaluation of the expectation value $\la\gamma_k^\dag\gamma_k\ra$ in this state, allows to compute the  exact kink number statistics using the  characteristic function in Eq. (\ref{charfunc}).
The comparison between analytical and numerical results is shown in a double logarithmic representation in Figure \ref{fig1} for the first few cumulants of the distribution ($q=1,2,3$) as a function of the quench rate $\tau_Q$.  The three cumulants are shown to exhibit a universal power-law scaling $\kappa_q\propto\tau_Q^{-1/2}$, consistent with the KZM prediction $\alpha=D\nu/(1+\nu z)=1/2$ for the 1D quantum Ising model with critical exponents $\nu=z=1$. Specifically,  a linear fit to the data in Fig. \ref{fig1}  for quench times $\tau_Q \in[2,200]$ yields the power-law exponents $\alpha=(0.503,0.507,0.539)$ 
for $q=1,2,3$ respectively. Figure \ref{fig1} also shows deviations from the scaling limit are first signaled by the third cumulant. The  $R^2$ coefficient for the fit to $\kappa_3$ is  $0.997$, in contrast with the unit value for $q=1,2$. 
The range of quench times in which the scaling limit holds decreases in high-order cumulants that are more sensitive to finite-size effects.
Despite the nonzero values of the latter, Figure  \ref{fig2} shows that the approximation of $P(n)$ by the  normal distribution 
$\mathbf{N}(Nd,3Nd/\pi^2)$ becomes highly accurate for slow quench rates, in the regime where  universal KZM power-law scaling holds. We note that the scaling with the quench rate  not only breaks down at fast quenches but also at the onset of adiabaticity when $\la n\ra<1$, i.e., $\tau_Q> \hbar N^2/(8\pi^2J)$. Further, we note that in this limit the kink statistics is not simply described by the corresponding truncated normal distribution. The power-law scaling can however be prolonged to larger values of $\tau_Q$ by increasing the system size $N$, as shown in \cite{SM}.

%
\begin{figure}[t]
\begin{center}
\includegraphics[width=1\linewidth]{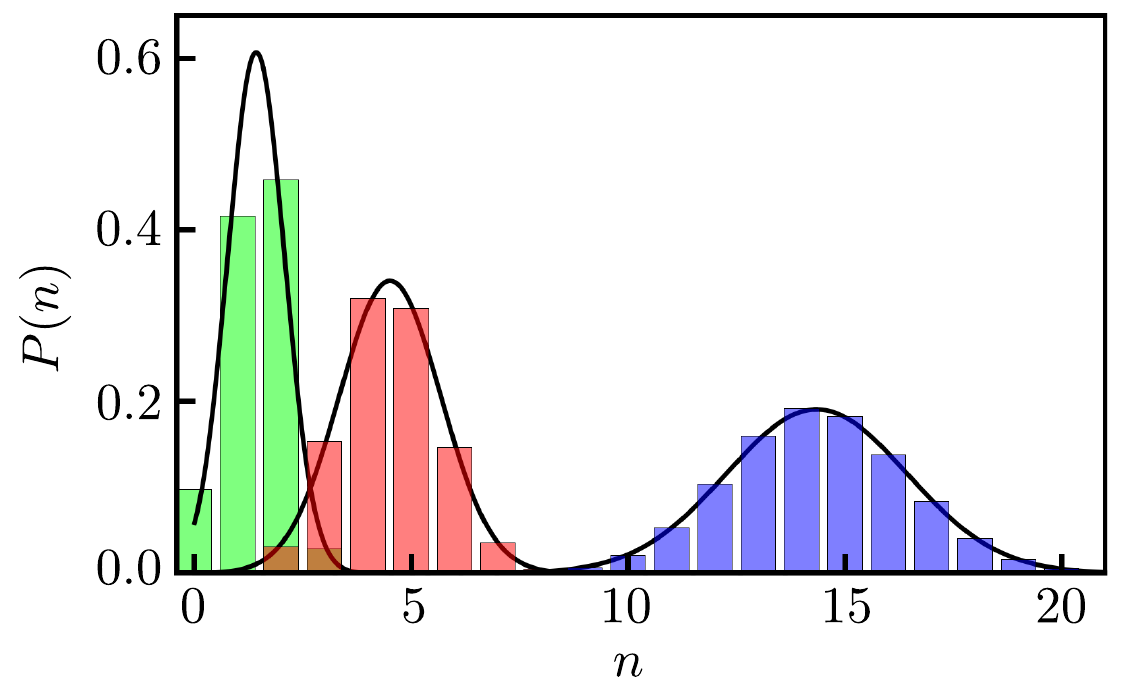}
\end{center}
\caption{\label{fig2}  {\bf Kink number distribution.} Dependence of the kink number distribution on the quench time $\tau_Q$ at which a 1D quantum Ising chain is driven through the quantum phase transition from the paramagnetic to the ferromagnetic phase. From right to left,  $\tau_Q=10,100,1000$  ($N=400$). In spite of the non-zero cumulants $\kappa_q$ with $q>2$ the distribution in the scaling limit is well approximated by the normal distribution in Eq. (\ref{Gausspn}) with   $\kappa_1=Nd$ and $\kappa_2=3\la n\ra/\pi^2$ (solid lines). However, deviations become apparent at the onset of the adiabatic dynamics.
}
\end{figure}

{\it Summary.---}
In a quantum phase transition, the closing of the gap leads to a divergence of the relaxation time, known as critical slowing down.
As a result, the dynamics across a quantum critical point is nonadiabatic and results in the formation of topological defects. The paradigmatic framework to describe their formation is the Kibble-Zurek mechanism, whose main prediction is the universal scaling of the mean defect density with the quench time. 
 We have investigated the full counting statistics of topological defects formed in a quantum Ising chain and shown that the kink number distribution inherits a universal dependence on the quench rate.  The kink statistics  is found to be described by the Poisson binomial distribution, that should be common to quasi-free fermion models. In particular, all cumulants are proportional to the mean and obey a  power-law scaling with the quench time, dictated by the critical exponents of the universality class to which the system belongs. When the number of domains is large, the kink statistics becomes normal (Gaussian distributed), a feature that is expected to hold broadly, whenever the Kibble-Zurek mechanism applies.  Thus, the formation of topological defects across a quantum phase transition exhibits a signature of universality that is not restricted to the mean value, predicted by the Kibble-Zurek mechanism, but extends to the full counting statistics. 
The universal dependence of the counting statistics on the quench time should find widespread applications in nonequilibrium statistical mechanics, quantum simulation, quantum annealing, and quantum error suppression algorithms. Further, it constitutes an experimentally testable prediction with current quantum technology. In particular, it is accessible via quantum simulation in various quantum platforms including superconducting qubits, Rydberg gases and trapped ions.

{\it Acknowledgment.-}  
It is a pleasure to thank Fabian Essler for early discussions as well as Aur\'elia Chenu, Fernando J. G\'omez-Ruiz and John Gough for comments on the manuscript. This work has been partially supported by Institut Henri Poincar\'e  and  CNRS via the thematic trimester of the Centre  \'Emile Borel  ``Measurement and control of quantum systems : theory and experiments'' in Spring 2018. 
Funding support from the John Templeton Foundation and UMass Boston (project P20150000029279) is further acknowledged.

\bibliography{fcs_kinks_Bib}	

\newpage
\pagebreak

\clearpage
\widetext

\subsection{Diagonalization of the Quantum Ising Chain}
 
The quantum Ising chain is an instance of a quasi-free fermion model. By a combination of a Jordan-Wigner transformation and the introduction of Fourier modes it can be mapped to an ensemble of independent two-level systems. The required steps have been presented in a number of texts \cite{Sachdev,Chakrabarti96} and are reviewed here for the convenience of the reader.

The Jordan-Wigner transformation maps spin operators $\sigma_n^\alpha$ to fermionic operators $c_n$ satisfying anti-commutation relations 
$\{c_n,c_m^\dag\}=\delta_{nm}$ together with $\{c_n,c_m\}=0$.
The Jordan-Wigner transformation  is highly nonlocal and reads 
\beqa
\sigma_n^x&=&1-2c_n^\dag c_n,\\
\sigma_n^z&=&-(c_n+c_n^\dag)\prod_{m<n}(1-2c_m^\dag c_m).
\eeqa
In the study of the Ising chain, it is convenient to introduce the parity operator $\Pi$ that has eigenvalues $\{-1,+1\}$  corresponding to the number of fermions being odd and even, respectively.
Using it, the Ising chain Hamiltonian can be written as the direct sum of its projection onto each subspace
\beqa
\mathcal{H}=P_+\mathcal{H}_+P_+ +P_-\mathcal{H}_{-}P_-
\eeqa
where $P_\pm=(1\pm \Pi)/2$, and the corresponding projections read
\beqa
\mathcal{H}_\pm/J=-Ng+\sum_n\left(2gc_n^\dag c_n-c_n c_{n+1}- c_{n+1}^\dag c_n -c_n^\dag c_{n+1}-c_n^\dag c_{n+1}^\dag\right),
\eeqa
with  the boundary conditions  $c_{N+1}=\mp c_1$ for $\mathcal{H}_\pm$. We consider $N$ to be even.
The direct sum structure of the Hamiltonian  $\mathcal{H}$ carries over the time-evolution operator $U(t,t')$ that thus preserves parity during the dynamics, $[\Pi,\mathcal{H}]=[\Pi,U(t,t')]=0$.
From here on, we  focus on the even parity subspace to which the ground state of $\mathcal{H}$ belongs. The Hamiltonian $\mathcal{H}_+$ can be simplified introducing the Fourier modes $c_k$,
\beqa
c_n=\frac{e^{-i\pi/4}}{\sqrt{N}}\sum_{k\in K}c_ke^{ikn}, 
\eeqa
where  the  lattice spacing is taken to be unity and the wavevector $k$ takes values over the discrete set
\beqa
K=\left\{\pm(2\ell-1)\frac{\pi}{N}\bigg\vert \ell=1,\dots,\frac{N}{2}\right\}.
\eeqa
In terms of $\{c_k\}$, the Hamiltonian $\mathcal{H}_+$ becomes
\beqa
\mathcal{H}_+/J&=&-Ng+\sum_k[2(g-\cos k)c_k^\dag c_k+\sin k(c_k^\dag c_{-k}^\dag +c_{-k}c_k)] \\
& =& -Ng+2\sum_{k>0} \psi_k^\dagger \left[ \sigma_k^z (g-\cos k)+\sigma_k^x \sin k  \right] \psi_k, \label{Hntls}
\eeqa
where  $\psi_k^\dagger \equiv (c_k^\dagger, c_{-k})$.
The time-dependent magnetic field term $-Ng$ simply contributes to a phase and can be gauged away.
Equation (\ref{Hntls}) is the representation of the Ising Hamiltonian used in the main text (where we drop the subscript $+$), and describes each $k$-mode as an independent two-level system.

\subsection{Power-Law Behavior  of Cumulants for Finite System Size}

%
\begin{figure}[t]
\begin{center}
\includegraphics[width=0.6\linewidth]{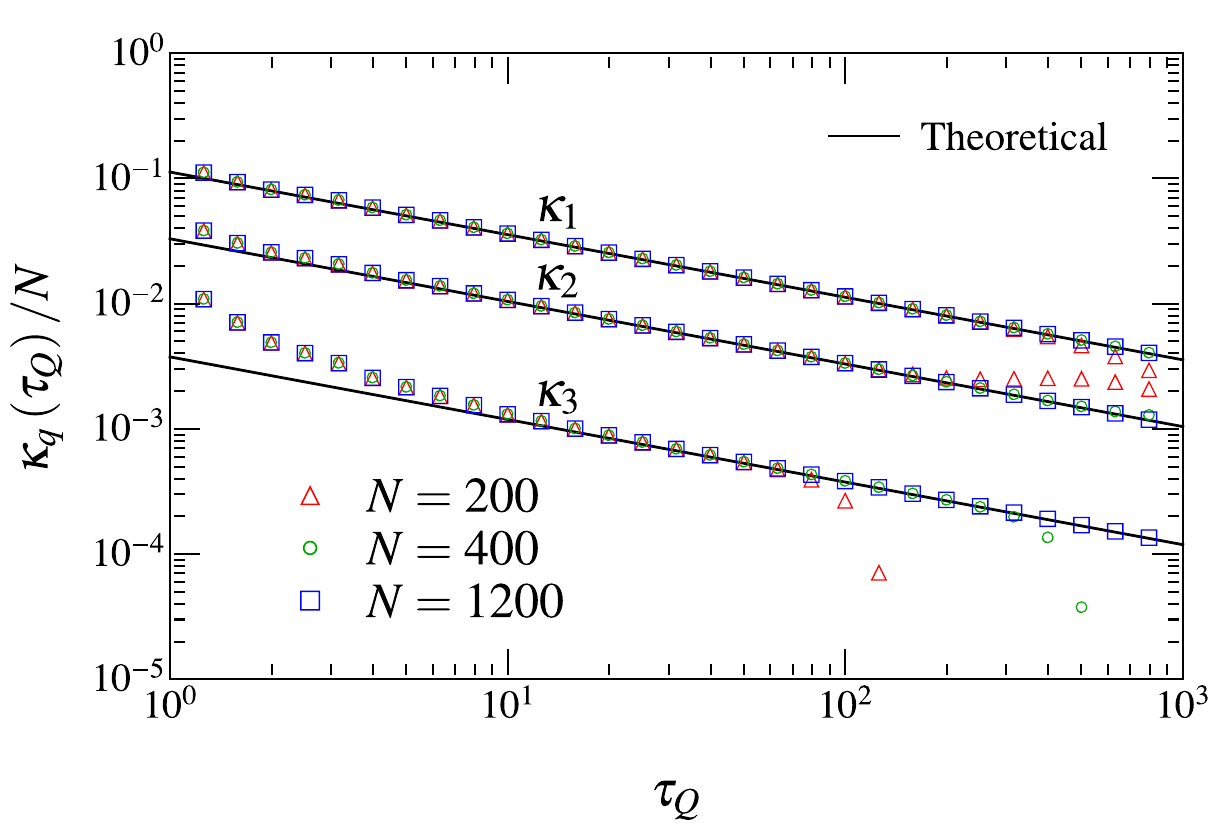}
\end{center}
\caption{\label{fig1SM}  {\bf Convergence of power-law scaling of cumulants $\kappa_q(\tau_Q)/N$ with increasing system size.}  From top to bottom, universal scaling of the mean density of defects ($q=1$), the corresponding variance ($q=2$) and the third cumulant ($q=3$) of the kink number distribution as a function of the quench time $\tau_Q$. The power-law behavior breaks down at fast quenches, away from the scaling regime. These deviations are largely independent of  the system size. The power-law scaling is also limited by the onset of adiabatic dynamics at slow quenches. These deviations can be suppressed by increasing the system size, thus prolonging the range of values in which the power-law scaling holds. Lines in black correspond to the analytical expression quoted in the main text in the scaling regime. }
\end{figure}

The analytical expression provided for the distribution of topological defects is essentially exact. It relies only on two approximations, the validity of which is well established.
First, the dynamics in each $k$-mode is described as a Landau-Zener crossing \cite{Dziarmaga05}. Second, the sum over the discrete set of modes is approximated by a an integral, $\sum_k\rightarrow\frac{N}{2\pi}\int_{-\pi}^\pi dk$. With these two approximations, we have derived a general expression valid for moderate quench times -- Eq. (14) in the main text -- and the scaling limit associated with an expansion to leading order in $1/\tau_Q$, i.e., Eq. (16) in the main text.

It should be clear that deviations from the power-law behavior governing the dependence of the cumulants $\kappa_q$ on the quench time $\tau_Q$ are predominantly associated with the consideration of moderate quench times away from the scaling limit and finite-size effects. The relevance of the later becomes apparent by studying different system sizes.
In particular, in the limit of slow quenches, the scaling law is expected to break down whenever the number of defects is small, e.g., $\langle n\rangle=\frac{N}{2\pi} \sqrt{\frac{\hbar}{2J\tau_Q}}\ll1 $, which leads to the adiabaticity condition
discussed in the text, $\tau_Q> \hbar N^2/(8\pi^2J)$. 
Figure \ref{fig1SM} shows the power-law scaling of the first three cumulants for  different values of the system size $N$. Deviations from the scaling regime are observed for fast quench times. Their existence  is  independent of the system size and they arise due to the contribution of the error function ${\rm erf}(x)$, e.g.,  in  Eq. (14) in the main text; see as well Eqs. (\ref{ekappa1})-(\ref{ekappa2}). 
By contrast, the breakdown of the power laws at slow quench times stems from the onset of adiabatic dynamics that can be postponed to higher values of $\tau_Q$ by increasing  $N$. In short, the regime of quench times in which the scaling law holds can  be extended  by increasing the system size $N$.

\subsection{ Non-Poissonian  Limit of the Kink Distribution}

A theorem on probability theory shows that the Poissonian binomial distribution can be approximated by a Poissonian distribution in the limit of large number of trials under certain conditions \cite{LeCam60}. We  next consider the application of this result to the dynamics of quantum phase transitions and show that the kink distribution can {\it not} be approximated by the Poisson distribution in the scaling regime. Said differently, the creation  of  kinks is not an independent process.

To this end,  let ${x_1,x_2,\dots,x_N}$ be the Bernouilli variables associated with the Fourier modes. Each Bernouilli variable $x_k$ takes outcomes $1$ and $0$ with probabilities $p_k$ and $(1-p_k)$, respectively. The probability for the $k$-mode to be found in the excited state is denoted by ${\rm Prob}(x_k=1)=p_k$, where $p_k$ can be computed by the Landau-Zener formula. We consider the sum of the excitation probabilities in each mode which equals the mean kink number $\sum_kp_k=\la \hat{\mathcal{N}}\ra=\kappa_1=\la n\ra$. We also introduce the variable $S=\sum_kx_k$ which is a classical variable equivalent to the kink number operator $\hat{\mathcal{N}}$.
Le Cam's theorem bounds the total variation between the probability distribution of $P(S=n)=P(n)$ and the Poisson distribution according to
\beqa
\sum_{n=0}^{N-1}\left|P(n)-\frac{\la n\ra^ne^{-\la n\ra}}{n!}\right|<2\sum_{k=1}^Np_k^2.
\label{ideq}
\eeqa
The term in the rhs can be evaluated using Landau-Zener's formula $p_k=\la \gamma_k^\dagger \gamma_k\ra=\exp\left(-\frac{1}{\hbar}2\pi J\tau_Qk^2\right)$ 
\beqa
2\sum_{k=1}^Np_k^2&=&4\sum_{k>0}\exp\left(-\frac{1}{\hbar}4\pi J\tau_Qk^2\right)\\
&=&4\sum_{\ell=1}^{N/2}\exp\left(-\frac{1}{\hbar}4\pi J\tau_Q \frac{\pi^2}{N^2}(2\ell-1)^2\right).
\eeqa
Indeed, in the continuum limit, 
\beqa
2\sum_kp_k^2&=&\frac{N}{\pi}\int_{-\pi}^\pi\exp\left(-\frac{1}{\hbar}4\pi J\tau_Qk^2\right)\\
&=&\sqrt{2}\la n\ra\times {\rm Erf}\left(2\pi^{3/2}\sqrt{J\tau_Q/\hbar}\right)\\
&\sim&\sqrt{2}\la n\ra,
\eeqa
where the last line holds in the limit of slow quenches.
As a result, the integrated difference between the kink distribution $P(n)$ and the Poissonian approximation, Eq. (\ref{ideq}), is always of the order of the mean number of kinks and diverges in the thermodynamic limit $N\rightarrow \infty$.
The fact that the kink distribution cannot be approximated by a Poisson distribution can also be inferred from the fact that for $P(n)$ the first cumulant never equals the second one and indeed $\kappa_2=(1-1/\sqrt{2})\kappa_1$. 

Therefore, the  formation of one kink is not independent from the formation of other kinks. In the Fourier description, even if the dynamics  in each $k$-mode is independent of the others,  excitation probabilities are correlated  between different modes due to the spectrum of the Ising chain and the applicability  the Landau Zener formula. Indeed, for two different modes with wavevectors $k_1$ and $k_2$ the corresponding excitation probabilities are related by $p_{k_1}/p_{k_2}=e^{-2\pi J\tau_q(k_1^2-k_2^2)/\hbar}$.

\subsection{Normal Limit of the Kink Distribution}
As shown in the main text, the kink distribution can be approximated by a normal distribution in the limit of a large number of spins $N$, far away from the onset of adiabatic dynamics. 
The power-series expansion of the cumulant generating function reads
\beqa
\label{eqCGF}
\log  \widetilde{P}(\theta;\tau_Q)=\sum_{q=1}^\infty \frac{(i\theta)^q}{q!}\kappa_q=i\theta \kappa_1-\frac{\theta^2}{2}\kappa_2+\dots
\eeqa
Clearly, the truncation to second order yields a normal distribution
\beqa
P(n)&=&\frac{1}{2\pi}\int_{-\pi}^{\pi}d\theta \widetilde{P}(\theta;\tau_Q) e^{-i\theta n}\\
&\simeq& \frac{1}{2\pi}\int_{-\pi}^{\pi}d\theta \exp\left(i\theta \kappa_1-\frac{\theta^2}{2}\kappa_2\right) e^{-i\theta n}\\
&=&\frac{1}{\sqrt{2\pi\kappa_2}}\exp\left[-\frac{(n-\kappa_1)^2}{2\kappa_2}\right].
\eeqa
Matching powers between the analytical expression
\beqa
\log  \widetilde{P}(\theta;\tau_Q)
=-\sum_{p=1}^\infty\frac{(1-e^{i\theta})^p}{p\sqrt{p}}N d\times {\rm erf}\left(\frac{\sqrt{\pi p}}{2d}\right),
\eeqa
and Eq. (\ref{eqCGF}) one finds the  first two cumulants to be given by
\beqa
\label{ekappa1}
\kappa_1&=&\frac{N}{2\pi}\sqrt{\frac{\hbar}{2J\tau_Q}}\,{\rm erf}\left(\pi\sqrt{\frac{2\pi J \tau_Q}{\hbar}}\right),\\
\kappa_2&=&\frac{N}{2\pi}\sqrt{\frac{\hbar}{2J\tau_Q}}\left[{\rm erf}\left(\sqrt{\frac{2\pi^3 J \tau_Q}{\hbar}}\right)-\frac{1}{\sqrt{2}}{\rm erf}\left(\sqrt{\frac{4\pi^3 J \tau_Q}{\hbar}}\right)\right].\label{ekappa2}
\eeqa
The error functions rapidly approach the unit value whenever its argument exceeds $\sim 2$.
For large quench times, one thus obtains the expressions in the scaling limit quoted in the text, 
\beqa
\kappa_1&=&\frac{N}{2\pi}\sqrt{\frac{\hbar}{2J\tau_Q}},\\
\kappa_2&=&\frac{N}{2\pi}\sqrt{\frac{\hbar}{2J\tau_Q}}\left(1-\frac{1}{\sqrt{2}}\right).
\eeqa
The third cumulant $\kappa_3$ is roughly an order of magnitude smaller than $\kappa_2$, and thus  the normal distribution constitutes a relevant approximation to the  the exact kink number distribution. We note, however, that the ratio between cumulants is  essentially constant in the scaling limit and independent of the system size $N$. Therefore,  the kink distribution exhibits a nonzero skewness $\gamma_1\equiv \kappa_3/\kappa_2^{3/2}\approx 0.135$, even in the thermodynamic limit when $N\rightarrow\infty$.
\subsection{High-order cumulants}
For completeness, we note that an arbitrary cumulant $\kappa_q$ of the Poisson binomial distribution can be written as \cite{Lee95}
\beqa
\kappa_q=\sum_k\left[p(1-p)\frac{d}{dp}\right]^{q-1}p\bigg|_{p=p_k}
\eeqa
For the kink distribution of the quantum Ising chain one thus finds
\beqa
\kappa_q=\frac{N}{2\pi}\int_{-\pi}^\pi dk\left[p(1-p)\frac{d}{dp}\right]^{q-1}p\bigg|_{p=e^{-\frac{2\pi J\tau_Q}{\hbar}k^2}}.
\eeqa
The result expressions are somewhat cumbersome and will not be listed here.
For large quench times,  the lower and upper limit of the integral can be extended to $-\infty$ and $\infty$, and the corresponding expression of the cumulant in the scaling limit is found.
This is equivalent to make a $1/\tau_Q$ series expansion of the exact results, derived from the definite integral with upper and lower limits $\pm \pi$.
The remaining first ten cumulants in the scaling limit read
\beqa
\kappa_3/\la n\ra&=&1-\frac{3}{\sqrt{2}}+\frac{2}{\sqrt{3}}\approx 0.033,\\
\kappa_4/\la n\ra&=&-2-\frac{7}{\sqrt{2}}+4\sqrt{3}\approx -0.02154,\\
\kappa_5/\la n\ra&=&-29-\frac{15}{\sqrt{2}}+\frac{50}{\sqrt{3}}+\frac{24}{\sqrt{5}}\approx -0.005962,\\
\kappa_6/\la n\ra&=&-194-\frac{31}{\sqrt{2}}+60 \sqrt{3}+72 \sqrt{5}-20
   \sqrt{6}\approx 0.009838,\\
\kappa_7/\la n\ra&=&-1049-\frac{63}{\sqrt{2}}+\frac{602}{\sqrt{3}}+672 \sqrt{5}-420
   \sqrt{6}+\frac{720}{\sqrt{7}}\approx 0.003544,\\
\kappa_8/\la n\ra&=&-5102-\frac{2647}{\sqrt{2}}+644 \sqrt{3}+5040 \sqrt{5}-5320 \sqrt{6}+2880
   \sqrt{7}\approx -0.009979,\\
\kappa_9/\la n\ra&=&-9869-\frac{90975}{\sqrt{2}}+\frac{6050}{\sqrt{3}}+\frac{166824}{\sqrt{5}}-52920 \sqrt{6}+47520
   \sqrt{7}\approx -0.004320,\\
\kappa_{10}/\la n\ra&=&502486-\frac{1890511}{\sqrt{2}}+6220 \sqrt{3}+204120 \sqrt{5}-456540 \sqrt{6}+604800 \sqrt{7}-36288
   \sqrt{10}\approx  0.01761.
\eeqa

\end{document}